\documentclass[12pt]{article}

\pdfoutput=1

\usepackage{placeins}
\usepackage{amsmath,amssymb,amsfonts,amsthm,bm}    % Typical maths resource packages
\usepackage{algorithm2e}
\usepackage{graphicx}
\usepackage{lineno}
\usepackage{pdfpages}
\usepackage[onehalfspacing]{setspace}
\usepackage[margin=1.5in]{geometry}
\usepackage{setspace}
\usepackage{placeins}
\usepackage{url}
\usepackage{xcolor}
\usepackage{mathtools}
\usepackage{titlesec}
\usepackage{natbib}
\usepackage{fixmath}
\usepackage{bbm}
\usepackage{booktabs}
\usepackage{comment}
\usepackage{commath}
\usepackage{booktabs}
\usepackage{multirow,bigdelim}
\usepackage{subfigure}
\usepackage{authblk}

\titlelabel{\thetitle.\quad}

\newcommand{\doa}{May 14, 2020}

\begin{document}
	
\def\spacingset#1{\renewcommand{\baselinestretch}%
{#1}\small\normalsize} \spacingset{1}
	
%%%%%%%%%%%%%%%%%%%%%%%%%%%%%%%%%%%%%%%%%%%%%%%%%%%%%%%%%%%%%%%%%%%%%%%%%%%%%%

\title{\bf Nowcasting fatal COVID-19 infections  on a  regional level in Germany}
\author{Marc Schneble\textsuperscript{1}, Giacomo De Nicola\textsuperscript{1}, G\"oran Kauermann\textsuperscript{1} \\ and Ursula Berger\textsuperscript{2}}
\affil{\textsuperscript{1} Department of Statistics, Ludwig-Maximilians-University Munich} 
\affil{\textsuperscript{2} Institute for Medical Information Processing, Biometry, and Epidemiology, Ludwig-Maximilians-University Munich}

\date{}	
		
\maketitle

\begin{center}
\textbf{ - preliminary version, submitted to Biometrical Journal  -} \\
Please refer to and cite the following published article in Biometrical Journal \\ 
Schneble M, De Nicola G, Kauermann G, Berger U. Nowcasting fatal COVID-19 infections on a regional level in Germany. Biometrical Journal. 2020;1–19. \url{https://doi.org/10.1002/bimj.202000143}
\end{center}
	
\bigskip
\begin{abstract}
\noindent 
We analyse the temporal and regional structure in mortality rates related to COVID-19 infections. We relate the fatality date of each deceased patient to the corresponding day of registration of the infection, leading to a nowcasting model which allows us to estimate the number of present-day infections that will, at a later date, prove to be fatal. The numbers are broken down to the district level in Germany. Given that death counts generally provide more reliable information on the spread of the disease compared to infection counts, which inevitably depend on testing strategy and capacity, the proposed model and the presented results allow to obtain reliable insight into the current state of the pandemic in Germany.
\end{abstract}
	
\noindent%

{\it Keywords: Nowcasting; COVID-19; Generalized regression model; Disease mapping.}

\newpage

\section{Introduction}
In March 2020, COVID-19 became a global pandemic. 
From Wuhan, China, the virus spread across the whole world,
and with its diffusion, more and more data became
available to scientists for analytical purposes. In daily reports, 
the WHO provides the number of registered infections as well as the daily death
toll globally (\texttt{https://www.who.int/}). It is inevitable for
the number of registered infections to depend on the testing
strategy in each country (see e.g.\ \citealp{Cohen:2020}). 
This has a direct influence on the number of undetected infections
(see e.g.\ \citealp{Li-etal:2020}), and first empirical analyses aim
to quantify how detected and undetected infections are
related (see e.g.\ \citealp{Niehus-etal:2020}).
Though similar issues with respect to data quality hold
for the reported number of fatalities (see
e.g.\ \citealp{Baud-etal:2020}), the number of deaths can overall be considered a more reliable source of
information  than the number of registered infections. The results of the "Heinsberg Study" in Germany point in the same direction \citep{heinsberg:20}.
A thorough analysis of death counts can in turn generate insights on changes in infections  as proposed in \citet{Flaxman-etal:2020} (see also \citealp{Fergueson-etal:2020}).
%The main focus of the first paper  was to quantify the effect of different non-pharmacological interventions, such as school closure or total lockdown, based on European data.
In this paper we pursue the idea of directly modelling 
registered death counts instead of registered infections. We analyse data from Germany and break down the
analyses to  a regional level. Such regional view is apparently immensely important, considering the local nature of some of the  outbreaks for example in Italy
(see e.g.\ \citealp{Grasselli-etal:2020a}, \citealp{Grasselli-etal:2020b}), France (see e.g.\ \citealp{Massonnaud-etal:2020}) or Spain.

The analysis of fatalities has, however, an inevitable time delay, and requires to take the course of the disease into account.
A first approach on modelling and analysing the time
from illness and onset of symptoms to reporting and further to death is given in \citet{Jung-etal:2020}
(see also \citealp{Linton-etal:2020}). Understanding the delay between  onset and
registration of an infection and, for  severe cases, the time between registered
infection and death can be of vital importance. 
Knowledge on those time spans allows us to
obtain estimates for the number of infections that are expected to be fatal based on the
number of infections registered on the present day. The statistical
technique to obtain such  estimates is called nowcasting (see e.g.\ \citealp{Hoehle-Heiden:2014}) and traces back to 
\citet{Lawless:1994}. 
Nowcasting in Covid-19
data analyses is not novel and is for instance used in  \citet{Guenther-etal:2020} for nowcasting daily infection counts, that is to adjust daily reported new infections to include infections which occurred the same day but were not yet reported. 
We extend this approach to model the delay between the registration date of an infection
and its fatal outcome.

% \textcolor{red}{This enables us to nowcast the number of fatal infections based on present-day death counts. (Wiederholung)}
%Assuming a constant case fatality rate, i.e. a stable proportion of death compared to the true number of infections when adjusting for age and gender, a differential analysis of the number of current fatal infections on a regional level allows to draw conclusions on current dynamics of the disease. 

We therefore analyse the number of fatal cases of Covid-19 infections in Germany using district-level data.
The data are provided by the Robert-Koch-Institute (\texttt{www.rki.de})  and give the cumulative number of deaths in different gender and age
groups for each of the 412 administrative districts in Germany together with the date of registration of the infection. The
data are available in dynamic form through daily downloads of the
updated cumulated numbers of deaths.
We employ flexible statistical models with smooth components (see
e.g.\ \citealp{Wood:2017}) assuming a district specific Poisson process.
% We employ flexible statistical models with smooth components, where the latter are estimated with standard smoothing %techniques (see e.g.\ \citealp{Wood:2017}). We assume a district specific Poisson process and focus on its resulting intensity.
The spatial structure in the  death rate is incorporated in two ways.
First, we assume a spatial correlation of the number of deaths by including a long-range smooth spatial death intensity.
This allows to show that regions of Germany are affected to different extents.
On top of this long-range effect we include two types of unstructured region specific
effects. An overall region specific effect reflects the situation of a district as a whole, while a short-term effect mirrors region specific variation of fatalities over time and captures local outbreaks as happened in e.g.\ Heinsberg (North-Rhine-Westphalia) or Tirschenreuth (Bavaria).
%On top of this long-range effect we include region specific effects, which capture local outbreaks as happened in e.g.\ %Heinsberg (North-Rhine-Westfalia) or Tirschenreuth (Bavaria). The region specific effects are unstructured, meaning that there %is no spatial correlation. Hence, single districts can have an increased death rate while neighbouring districts are less affected. 
In addition we include dynamic effects to capture the global changes in the number of fatal infections for Germany over calendar time. 
This enables us to investigate the impact of certain interventions, such as
social distancing, school closure, complete lockdowns and lockdown releases, on the dynamic of the infection and hence on the number of deaths.

Modelling infectious diseases is a well developed field in statistics
and we refer to \citet{Held-etal:2017} for a general overview of the
different models. We also refer to the powerful R package {\tt surveillance}
\citep{Meyer-etal:2017}.
%Generally, the Susceptible, Infectious, Removed (SIR) model builds
%upon an infection intensity, which may contain endemic as well as
%epidemic components.
%The endemic component describes general trends, while the epidemic part
%captures infectious behavior.
%We here concentrate on the endemic component only,  “which reflects sporadic events caused by unobserved sources
%of infection” (see \citealp{Meyer-etal:2017}).
Since our focus is on analysing the district specific dynamics of fatal infections we here make use of Poisson-based models implemented in the 
{\tt mgcv} package in \texttt{R}, which allows to decompose the spatial component in
more depth. 
%In particular, we are able to uncover sporadic region-specific
%events caused by unobserved local sources of infections.

The paper is organized as follows. In Section 2 we describe the
data. Section 3 highlights the results of our analysis. The remaining
sections provide the technical material, starting with
Section 4 where we motivate the statistical model, which is extended
by our nowcasting model in Section 5. Extended results as well as
model validation are given in Section 6, while Section 7 concludes the paper.

\section{Data}
We make use of the COVID-19 dataset provided by
the Robert-Koch-Institute for the 412 districts in Germany (which also include the twelve districts of Berlin separately). The data
are updated on a daily basis and can be downloaded
{from the Robert-Koch-Institute's website}. We have daily downloads of the data
for the time interval from March 27, 2020 until today. The subsequent
analysis was conducted on \doa, and was performed considering only deadly infections with registration dates from March 26, 2020 until May 13, 2020 (the day before the day of analysis).

The data contain the newly notified laboratory-confirmed COVID-19 infections and the
cumulated number of deaths related to COVID-19
for each district of Germany, classified by gender and age group.
Each data entry has a time stamp which corresponds to the
registration date of a confirmed Covid-19 infection. This means that the time stamp for a fatal outcome always refers to the registration date and \textit{not} to the death date. 
Due to daily downloads of the data we can derive the time point of death (or to be more specific, the time point when the death of a case is included in the database). 
We obtain the latter by observing a status change from
infected to deceased when comparing the data from two consecutive days.

The Robert-Koch-Institute collects the data from the district-based health authorities ({\sl Gesundheits{\"a}mter}).
Due to different population sizes in the districts and certainly also because
of different local situations, some health authorities report the daily numbers
to the Robert-Koch-Institute with a delay. This happens in particular over the weekend, a fact that we need to take
into account in our model. 

We refrain from providing general descriptive statistics of the data here, since these numbers can easily be found  on the RKI webpage, which also gives a link to a dashboard  to visualize the data (see also 
{https://corona.stat.uni-muenchen.de/maps/})

\section{Results of the Analyses}

\begin{table}[t]
	\center
	\begin{tabular}{rlrrr}
		\toprule
		& & \multirow{2}{*}{\textbf{Effect (s.e.)}}&  \textbf{exp(Effect) /}\\
		& & & \textbf{Relative Risk} \\
		\midrule
		\midrule
		& Intercept & -16.103 (0.079) & $1.02\cdot10^{-7}$\\
		\midrule
		\ldelim\{{4}{4.15cm}[Patient related effects] & Age 15-34 & -2.572 (0.260)  & 0.076\\
		& Age 60-79 & 2.261 (0.061) & 13.645\\
		& Age 80+ & 4.645 (0.059) & 104.101\\
		& Female & -0.503 (0.022) & 0.605\\
		\midrule
		\ldelim\{{6}{4.6cm}[Reporting related effects] & Tuesday & 0.188 (0.042) & 1.207\\
		& Wednesday & 0.241 (0.042) & 1.272\\
		& Thursday & 0.255 (0.041) & 1.291\\
		& Friday & 0.107 (0.042) & 1.113\\
		& Saturday & -0.128 (0.044) & 0.879 \\
		& Sunday & -0.406 (0.048) & 0.666\\
		\bottomrule
	\end{tabular}
	\caption{Estimated fixed linear effects (standard errors in brackets) in the quasi-Poisson-model. Parameters and their standard errors are given on the log scale. To facilitate interpretation, the multiplicative effect is also given (on the exp scale). The reference category for age is the age group 35-59. The reference group for the weekdays is Monday.}
	\label{tab: linear effects poisson}
\end{table} 

\begin{figure}[t]
	\center
	\includegraphics[width = 0.8\textwidth]{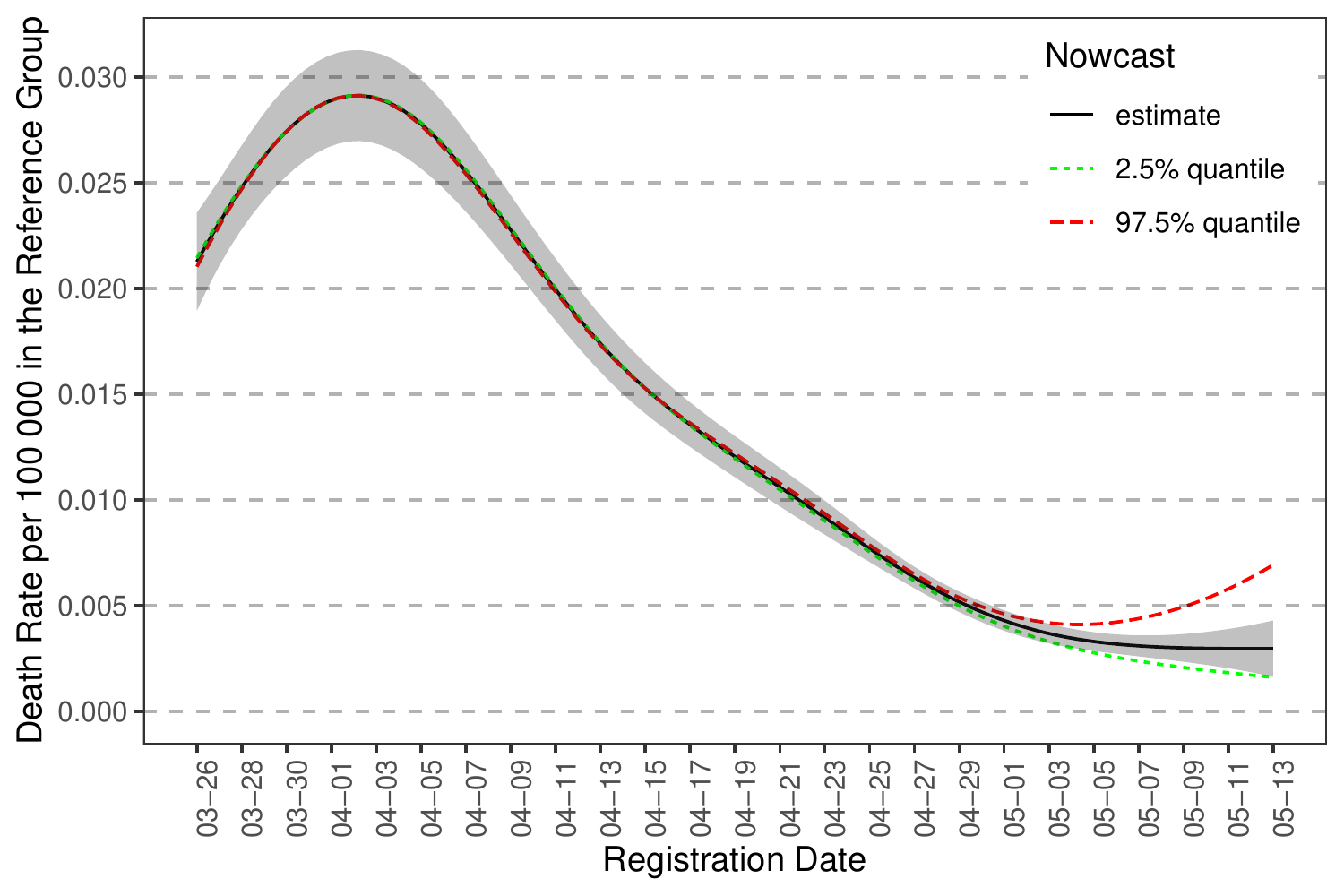} 
	\caption{Fitted smoothed death rate per 100 000 inhabitants in the reference group (males aged between 35 and 59 in an average district) including 95\% confidence bands as shaded area. Uncertainty resulting from the nowcast model is shown as dashed coloured lines. }
	\label{fig: death rate per 100k}
\end{figure}

\subsection{Fatal infections in Germany}  
Before we discuss our modelling approach in detail, we want to  describe our major findings. First, Table \ref{tab: linear effects poisson} shows that age and gender both play a major role when estimating the daily death toll. As is generally known, elderly people exhibit a much higher death rate which is for the age group 80+ around 100 times higher than for people in the age group 35-59. A remarkable difference is also observed between genders, where the expected death rate of females is around 40\% ($ \approx 1-\exp(-0.503$)) lower than the death rate for males.  Furthermore, we see that significantly less deaths are attributed to infections registered on Sundays compared to weekdays, due to the existing reporting delay during weekends.

\begin{figure}
	\center
	\includegraphics[width = \textwidth]{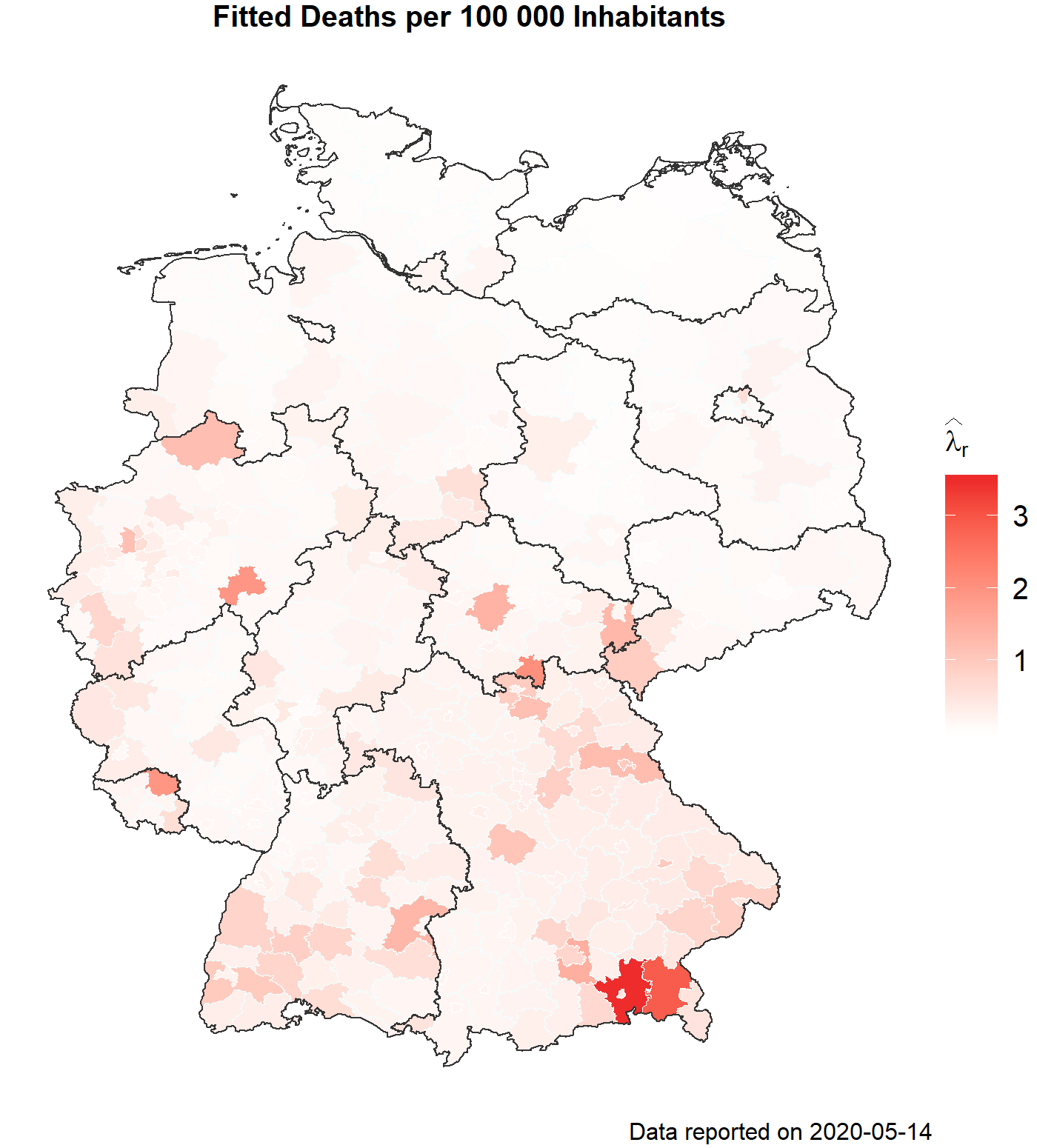} 
	\caption{Expected deaths per 100 000 inhabitants in each district in the week from Monday, May 4 until Sunday, May 10, 2020.}
	\label{fig: fited intensities}
\end{figure}

Our model includes a global smooth time trend representing changes in the death rate since March 26th. This is visualized in  Figure \ref{fig: death rate per 100k}. The plotted death rate is scaled to give the expected number of deaths per 100.000 people in an average district for the reference group, i.e. males in the age group 35 - 59. Overall, we see a peak in the death rate on April 3rd and a downwards slope till end of April. However, our nowcast reveals that the rate remains constant since beginning of May. Note that this recent development cannot be seen by simply displaying the raw death counts of these days.  The nowcasting step inevitably carries statistical uncertainty, which is taken into account in Figure \ref{fig: death rate per 100k} by including  best and worst case scenarios. The latter are based on  bootstrapped confidence intervals, where details are provided in Section \ref{sec:6.4} later in the paper. 

Our aim is to  investigate spatial variation and regional dynamics. To do so, we combine  a global geographic trend for Germany with  unstructured region-specific effects, where the latter  uncover local behaviour.  In Figure \ref{fig: fited intensities} we combine these different components and map the fitted nowcasted death counts related to Covid-19 for the different districts of Germany, cumulating over the last seven days before the day of analysis (here \doa).
While in most districts of Germany the death rate is relatively low, some hotspots can be identified.
Among those, Traunstein and Rosenheim (in the south-east part of Bavaria) are the most evident, but  Greiz and Sonneberg (east and south part of Thuringia) stand out as well, to mention a few. 
A deeper investigation of the spatial structure is provided in Section 6, where we show the global geographic trend and provide maps that allow to detect new hotspot areas, after correcting for the overall spatial distribution of the infection.

\subsection{Nowcasting the number of deaths}  
\begin{figure}[t]
	\center
	\includegraphics[width = \textwidth]{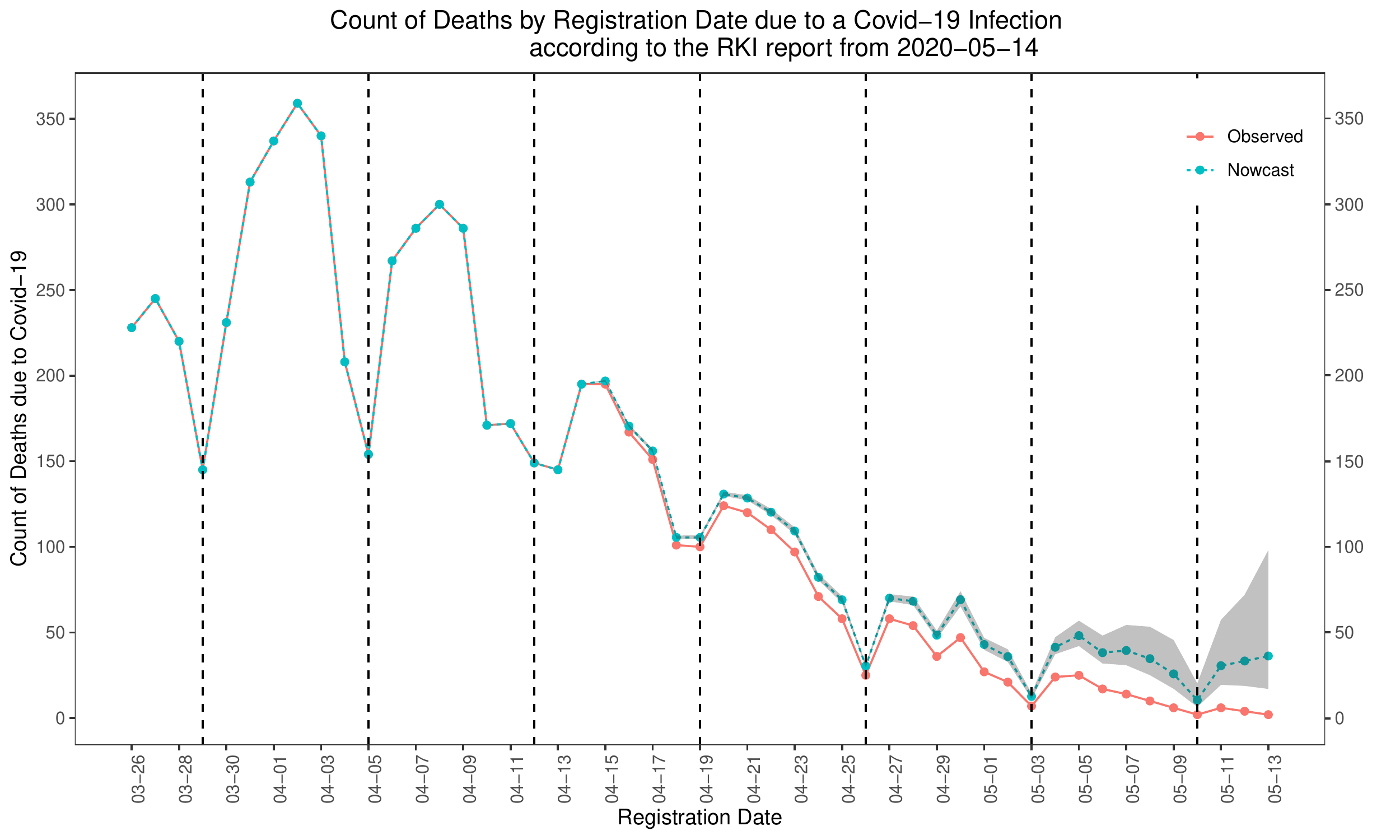}
	\caption{Nowcasting of daily death counts due to a Covid-19 infection including 95\% prediction intervals (shaded areas). Sundays are marked by a dashed vertical line.}
	\label{fig: nowcasting}
\end{figure}

On the day of analysis, we do not observe the total counts of deaths for recently registered infections, since not all patients with an ongoing fatal infections have died yet. We therefore nowcast those numbers, i.e.\ we predict the prospective deaths which can be attributed to all registration dates up to today. This is done on a national level, and the resulting nowcast  of fatal infections for Germany is shown in Figure \ref{fig: nowcasting}.
For example, on \doa{} there are 25 deaths reported where the infection was registered on May 5th (red line on May 5th). We expect this number to increase to about 50  when all deaths due to Covid-19 for this registration date will have been reported (blue line on May 5th). Naturally, the closer a date is to the present, the larger the uncertainty in the nowcast. This is shown  by the shaded bands. Details on how the statistical uncertainty has been quantified are provided in Section \ref{sec: nowcasting} below. The fit of this model has been incorporated into the  district model discussed before,  but the nowcast results are interesting in their own right. The curve confirms that the number of fatal infections is decreasing since the beginning of April. 
Note that the curve also mirrors the "weekend effect" in registration, as less infections are reported on Sundays. Further analyses and a detailed description of the model are given in the following sections. 

\section{Mortality Model}

Let $Y_{t,r,g}$ denote the number of daily deaths due to COVID-19 in
district/region $r$ and  age and gender group $g$ with time point (date of registration)
$t = 0,\dots,T$. Here $t = T$ corresponds to the day of analysis, which is \doa{}  and $t = 0$ corresponds to 
March 26, 2020. Note that time point $t$ refers to the time point of registration,
i.e. the date at which the infection was confirmed.  Even though the time point of infection obviously precedes that of death, registration can also occur after death, e.g.\ when a post mortem test is conducted, or when test results arrive after the patient has passed away. 
We set the day of death to be equal to the day of registered infections in this case.
The majority of fatalities
with registered infection at time point $t$ have not yet been observed at time $t$, as these
deaths will occur later. 
%Only patients which are registered as infected on their day of death are already available at time point $t$. 
We therefore need a model for nowcasting, which is discussed in the next section. For now we assume all $Y_{t,r,g}$ to be known.

We model $Y_{t,r,g}$ as (quasi-)Poisson distributed according to
\begin{equation}
\label{eq:modquasi}
Y_{t,r,g} \sim \mbox{(quasi-)}Poisson( \lambda_{t,r,g} )
\end{equation}
where we specify $\lambda_{t,r,g}$ through
\begin{align}
\begin{split}
\lambda_{t,r,g} &= \exp\{(\beta_0 + age_g \beta_{age} + gender_g\beta_{gender} + weekday_t \beta_{weekday} \\
& +  m_1(t) + m_2(s_r) + u_{r0}  +  1_{\lbrace t\geq T-14 \rbrace}u_{r1} +  \log(pop_{r,g})\}.
\label{eq: lambda rtg}
\end{split}
\end{align}
The linear predictor is composed as follows:
\begin{itemize}
	\item $\beta_0$ is the intercept.
	\item $\beta_{age}$ and $\beta_{gender}$ are the age and gender related regression coefficients.
	\item $\beta_{weekday}$ are the weekday-related regression coefficients.
	\item $m_1(t)$ is an overall smooth time trend, with no prior structure imposed on it.
	\item $m_2(s_r)$ is a smooth spatial effect, where $s_r$ is the geographical centroid of district/region $r$.
	\item $u_{r0}$ and $u_{r1}$ are district/region-specific random effects which are $i.i.d.$ and follow a Normal prior probability model. While $u_{r0}$ specifies an overall level of in the death rate for district $r$ over the entire observation time, $ u_{r1}$  reveals region specific dynamics by allowing the regional effects to differ for the last 14 days.
	\item $pop_{r,g}$ is the gender and age group-specific population size in district/region  $r$ and serves as an offset in our model.
\end{itemize}
We here emphasize that we fit two spatial effects of different types: We model a smooth spatial
effect, i.e.\ $m_2(s_r)$, which takes the correlation between the
death rates of neighbouring districts/regions into account and gives a global overview of the spatial distribution of fatal infections. In addition to that we also have unstructured district/region-specific effects $\boldsymbol{u}_r = (u_{r0}, u_{r1})^\top$, which capture local behaviour related to single districts only. The district specific effects $\boldsymbol{u}_r$ are
considered as random with a prior structure
\begin{equation}
\boldsymbol{u}_r \sim N(\boldsymbol{0}, \boldsymbol{\Sigma}_u) \mbox{ i.i.d}
\label{eq:u_r}
\end{equation}
for $r = 1,\dots,412$. The prior variance matrix $\boldsymbol{\Sigma}_u$ is estimated from the data. 
The predicted values $\widehat{\boldsymbol{u}}_r$ (i.e. the posterior mode) exhibit districts that show unexpectedly high or low death tolls when adjusted for the global spatial structure and for age- and gender-specific population size.

Model (\ref{eq:modquasi}) belongs to the model class of generalized additive mixed model, see e.g.\
\citet{Wood:2017}. The smooth functions are estimated by penalized splines, where the quadratic penalty can be comprehended as a Normal prior  (see e.g.\ \citealp{Wand:2003}). The same type of prior structure holds for the region-specific random effects $\boldsymbol{u}_r$. In other words, smooth estimation and random effect estimation can be accommodated in one fitting routine, which is implemented in the R package {\tt mgcv}. This package has been used to fit the model, so that no extra   software implementation was necessary.  This demonstrates the practicability of the method.

\section{Nowcasting Model}
\label{sec: nowcasting}
\subsection{Model Description}
The above model cannot be fitted directly to the available data, since we need to take 
the course of the disease into account. For a given registration date $t$, the number of deaths of patients registered as positive on that day, $Y_{t,r,g}$, may not yet be known, 
since not all patients with a fatal outcome of the disease have died yet.  
This requires the implementation of nowcasting.  
We do this on a national level, and cumulate the numbers over district/region $r$ and  gender and age groups $g$. 
This allows to drop the corresponding subscripts  in the following and we simply notate the
cumulated number of deaths with registered infections at day $t$ with $Y_t$.
%The majority of deaths occur with a delay in respect to the day of registered infection.
Let $N_{t,d}$ denote the number of deaths reported on day $t+d$ for infections registered on day $t$. Assuming that the true date of death is at $t+d$, or at least close to it, we ignore any time delays between time of death and its notification to the health authorities.
We call $d$ the duration between the registration date as a Covid-19 patient and the reported day of death, where $d
= 1, \ldots ,d_{max}$. Here, $d_{max}$ is a fixed reasonable maximum duration,
which we set to 30 days (see e.g.\ \citealp{Wilson:2020}). 
%Since the daily reported deaths of the RKI can at most be attributed to registration dates equal to the previous day, 
The minimum delay is one day.  
In nowcasting we are interested in the cumulated number of deaths for infections registered on day $t$, which we define as
\begin{equation*}
Y_t = \sum_{d=1}^{d_{max}} N_{t,d}.
\end{equation*}
The total number of deaths with a registered infection at $t$ is apparently unknown at time point $t$  and becomes available only after $d_{max}$ days. In other words, only after $d_{max}$ days we know exactly how many deaths occurred due to an infection which was registered on day $t$. We define the partial cumulated sum of deaths as
\begin{equation*}
C_{t,d} = \sum_{l=1}^d N_{t,l}
\end{equation*}
so that by definition $C_{t,d_{max}} = Y_t$.

On day $t = T$, when the nowcasting is performed, we are faced with
the following data constellation, where NA stands for not (yet) available:

\begin{center}
	\resizebox{\textwidth}{!}{\begin{tabular}{c|cccc|c}
			& \multicolumn{4}{c|}{d} & reported  \\
			t & 1 & 2&  $\cdots$ &   $d_{max}$ &  deaths \\
			\hline
			0 & $N_{0,1}$ & $N_{0,2}$ & $\cdots$ & $N_{0,d_{max}} $& $Y_0 $ \\
			1 & $N_{1,1}$ & $N_{1,2}$ & $\cdots$ & $N_{1,d_{max}} $& $Y_1 $ \\
			$\vdots$ &  $\vdots$ &  $\vdots$& $\vdots$ & $\vdots$ & $\vdots$ \\
			$ T-d_{max}$   & $ N_{T-d_{max},1}$ &  $ N_{T-d_{max},2}$ & $\cdots$ & $N_{T-d_{max}, d_{max}}$ & $Y_{T-d_{max}}$\\
			$ T-d_{max}+1$ & $N_{T-d_{max}+1,1}$ & $N_{T-d_{max}+1,2}$ & $\cdots$ & $\mbox{NA}$ & $C_{T-d_{max}-1, d_{max}-1}$\\
			$ \vdots$ &  $\vdots$ &   $\vdots$ & $\vdots$ & $\vdots$ & $\vdots$\\
			$ T-2 $& $N_{T-2,1}$ & $N_{T-2,2}$& $\mbox{NA}$  & $\mbox{NA}$ & $C_{T-2,2}$ \\
			$ T-1 $& $N_{T-1,1}$ &$\mbox{NA}$ & $\mbox{NA}$ & $\mbox{NA}$ & $C_{T-1,1}$
	\end{tabular}}
\end{center}

\vspace{0.5cm}

We may consider the time span between registered infection and (reported) death as a discrete duration time taking
values $d= 1, \ldots ,d_{max}$.
Let $D$ be the random duration time, which by construction is a  multinomial random variable.
In principle,
for each death we can consider the pairs $(D_i, t_i)$ as {\sl i.i.d.}
and we aim to find a suitable regression model for $D_i$ given $t_i$,
including potential additional covariates $x_{t,d}$.
We make use of the sequential multinomial model (see \citealp{Agresti:2010}) and define
\begin{equation*}
\pi(d;t,x_{t,d}) = P(D=d| D\le d;t,x_{t,d})
\end{equation*}
Let $F_t(d)$ denote the corresponding cumulated distribution function
of $D$
which relates to probabilities $\pi()$ through

\begin{align}
\begin{split}
F_t(d) & =  \mbox{P}_t(D \le d)  =  \mbox{P}(D\le d|D \le d+1) \cdot P(D \le d+1) \\
& =  (1-\pi(d+1;\cdot)) \cdot (1-\pi(d+2;\cdot)) \cdot \ldots \cdot (1-\pi(d_{max};\cdot)) \\
&= \prod_{k = d+1}^{d_{\max}} (1- \pi(k; \cdot))
\label{eq: F_t(d) prod}
\end{split}
\end{align}
for $d = 1,\dots,d_{\max}-1$ and $F_t(d_{\max}) =1$.

\vspace{0.5cm}

The available data on cumulated death counts  allow us  to estimate the
conditional probabilities $\pi(d;)$ for $d = 2,\dots,d_{\max}$. In fact, the sequential
multinomial model allows to look at binary data such that
\begin{equation}
\label{eq:nowcast-20}
N_{t,d} \sim (\text{quasi-)}Binomial \left(C_{t,d}, \pi(d;t,x_{t,d})\right)
\end{equation}
with
\begin{equation}
\label{eq:nowcast-2}
\mbox{logit}(\pi(d;t,x_{t,d})) =
s_1(t) + s_2(d) + x_{t,d} \gamma
\end{equation}
where 
\begin{itemize}
	\item $s_1(t)$ is an overall smooth time trend over calendar days,
	\item $s_2(d)$ is a smooth duration effects, capturing the course of the disease,
	\item $x_{t,d}$ are covariates which may be time and duration specific.
\end{itemize}
Assuming that $D$, the duration  between a registered fatal infection and its reported death, is independent of the number of fatal Covid-19 infections, we  obtain the relationship
\begin{equation}
\label{eq:nowcast-1}
E( C_{t,d})  = F_t(d) \cdot E(Y_{t}).
\end{equation}
Note further that if we model $Y_t$ with a quasi-Poisson model as
presented in the previous chapter, we have no available observation 
$Y_t$ for time points $t>T-d_{max}$. Instead,
we have observed $C_{t,T-t}$,  which relates to the mean of $Y_t$ through
(\ref{eq:nowcast-1}). Including  therefore  $\log F_t(T-t)$ as additional offset in model (\ref{eq: lambda rtg}), allows to fit the model
as before, but with nowcasted deaths included. That means, instead of $\lambda_{t,r,g}$ as in \eqref{eq: lambda rtg}, the expected death rates are now parametrized by $\lambda_{t,r,g}^\star = \lambda_{t,r,g} \exp( \log F_t(T-t))$, where the latter multiplicative term is included as additional offset in the model.

\subsection{Results for Nowcasting }

\begin{table}[t]
	\center
	\begin{tabular}{lrr}
		\toprule
		\textbf{ } & \textbf{Effect (s.e.)} & \textbf{exp(Effect)} \\
		\midrule
		Intercept & -2.843  (0.052)  & 0.058 \\
		Tuesday   & 0.049   (0.069) &  1.050\\
		Wednesday & 0.123   (0.069)  &  1.132\\
		Thursday  & 0.233    (0.066) & 1.262\\
		Friday    & 0.238    (0.069) & 1.307\\
		Saturday  & 0.268    (0.073) & 1.307\\
		Sunday    & 0.220    (0.079) & 1.246\\
		\bottomrule
	\end{tabular}
	\caption{Fixed effects for weekday of the registration date of the infection from the nowcasting model.}
	\label{tab: effects inverse survival}
\end{table}

\begin{figure}
	\centering
	\vspace{0cm}
	\subfigure{Smooth effect of calendar time} \vspace{0.1cm}
		%\caption{Smooth effect of calendar time}
		\includegraphics[width =\textwidth]{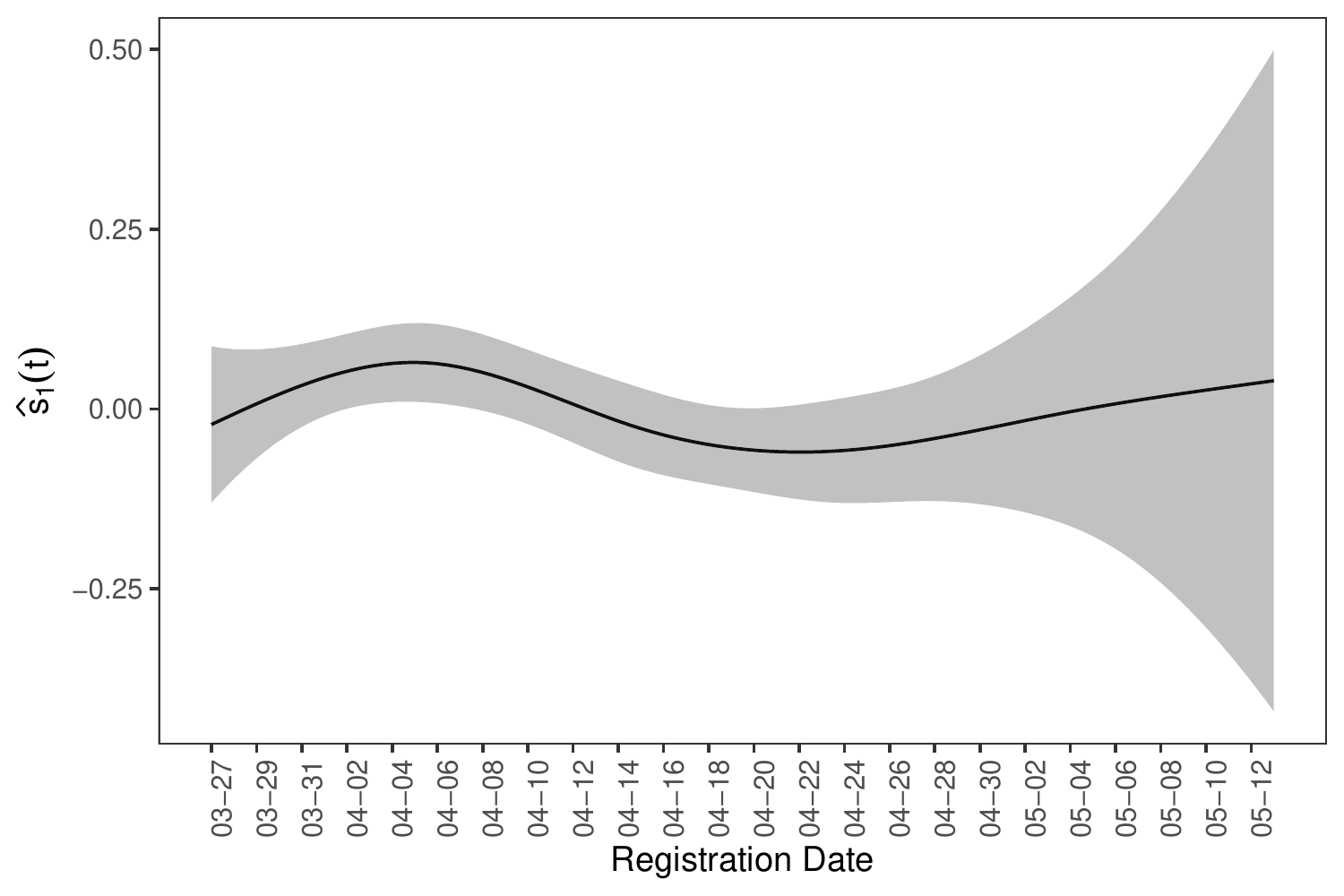}
	%\end{subfigure} \\
	\subfigure{Smooth effect of duration time}
		%\caption{Smooth effect of duration time}
		\includegraphics[width = \textwidth]{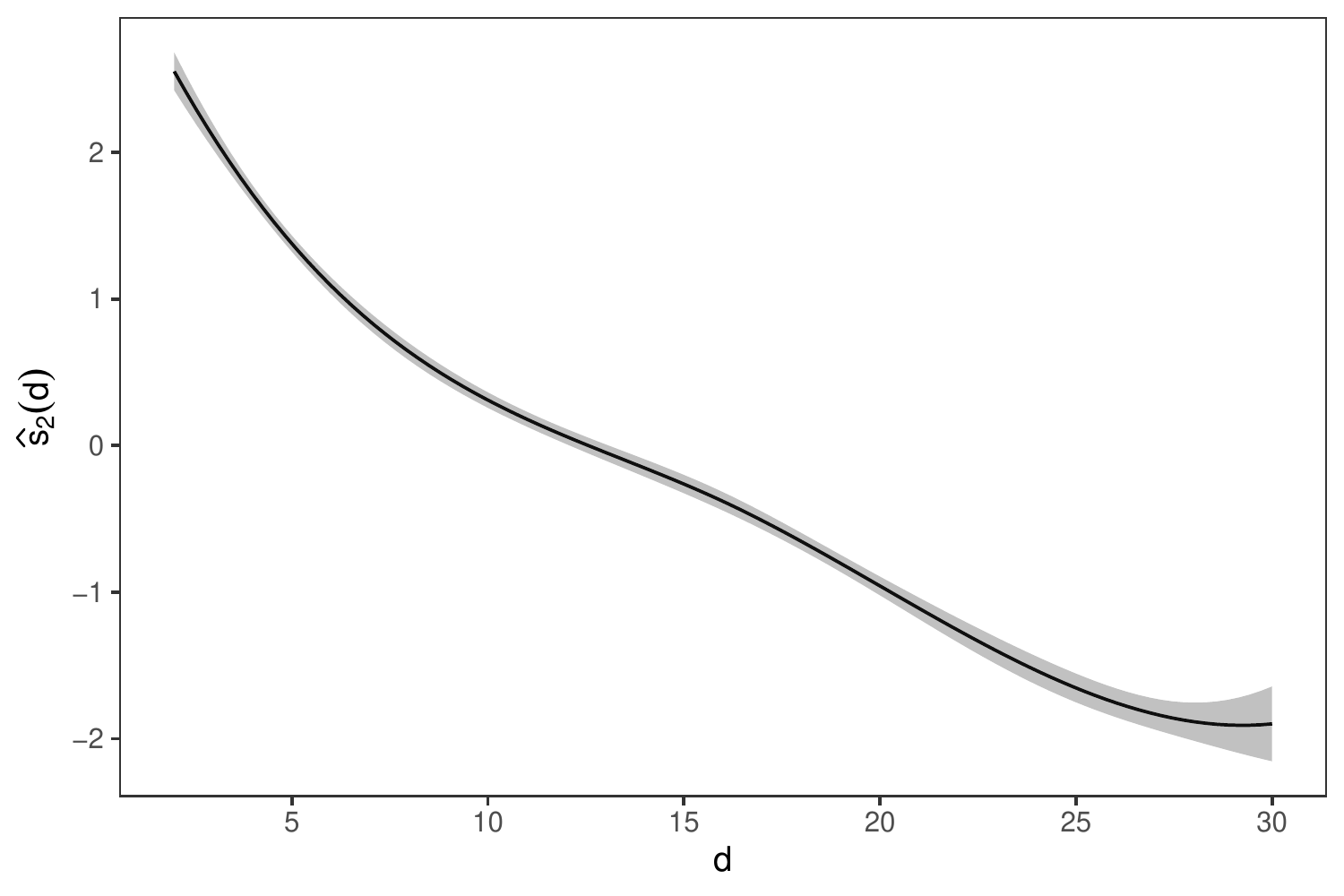}
		\vspace{-0.7cm}
	\caption{Estimates of smooth effects in the nowcasting model.}
	\label{fig:smooth effects nowcasting}
\end{figure}

%\begin{figure}
%	\center
%	\begin{subfigure}{0.8\textwidth}
%		\caption{Smooth effect of calendar time}
%		\includegraphics[width =\textwidth]{../../Plots/Nowcasting/TimeEffect/2020-05-14.pdf}
%		\label{fig: smooth time inverse survival}
%	\end{subfigure} \\
%	\begin{subfigure}{0.8\textwidth}
%		\caption{Smooth effect of duration time}
%		
%		\includegraphics[width = \textwidth]{../../Plots/Nowcasting/DurationEffect/2020-05-14.pdf}
%		\label{fig: smooth delay inverse survival}
%	\end{subfigure}
%	\caption{Estimates of smooth effects in the nowcasting model.}
%	\label{fig: smooth effects inverse survival}
%\end{figure}

We fit the nowcasting model (\ref{eq:nowcast-20}) with parametrization (\ref{eq:nowcast-2}).  We include a weekday effect for the registration date of the infection with reference category "Monday". The estimates of the fixed linear effects are shown in Table \ref{tab: effects inverse survival}.  The fitted smooth effects are shown in Figure \ref{fig:smooth effects nowcasting}, where the top panel shows the effect over calendar time, which is very weak and confirms that the course of the disease hardly varies over time.  This shows that the German health care system remained stable over the considered period, 
and hence survival did not depend on the date on which the infection was notified. 
%and hence the course of disease to death did never depend on the time point a patient's infection was registered. 

\begin{figure}
	\center
	\includegraphics[width = 0.8\textwidth]{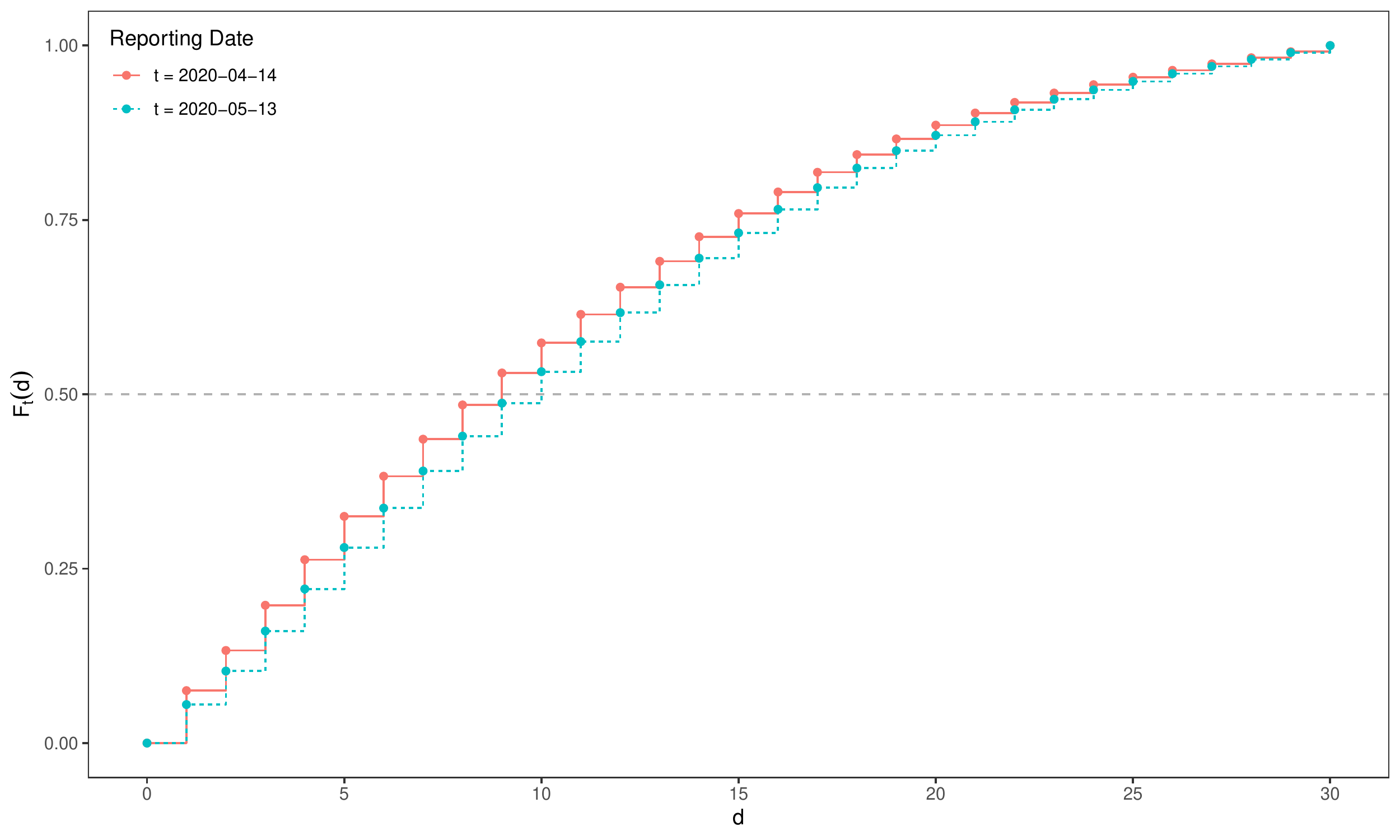} 
	\caption{Fitted distribution function $F_t(d)$ for two selected days: $t=$ Tuesday April 14, 2020 and $t = $ Wednesday May 13, 2020}.
	\label{fig: F_t(d} 
\end{figure}

The bottom panel of Figure \ref{fig:smooth effects nowcasting} shows the course of the disease as a smooth effect over the time between registration of the infection and death. We see that the probabilities $\pi(d; \cdot)$ decrease in $d$, where this effect is the strongest in the first days after registration. Thus, most of the Covid-19 patients with fatal infections are expected to die not long after their registration date. The effect of $d$ becomes easier to interpret by visualizing  the resulting distribution function $F_t(d)$. This  is shown in Figure \ref{fig: F_t(d} for two dates $t$, i.e.. April 14th and May 13th.  The plot also shows how the course of the disease hardly varies over calendar time: In fact, the small differences between the two distribution functions is dominated by the weekday effect, since the red curve is related to a Tuesday while the blue one is from a Wednesday.

\subsection{Uncertainty Quantification in Nowcasting }
\label{sec:bootstrap}
In Figure \ref{fig: nowcasting} above we have shown the nowcasting
results along with uncertainty intervals shaded in grey.
These  were constructed using a bootstrap approach as follows. 
Given the fitted model, we simulate $n = 10\text{ }000$ times from the asymptotic joint normal distribution of the estimated model parameters which results through \eqref{eq: F_t(d) prod}. This leads to a set of bootstrapped distribution functions
$\mathcal{F} = \lbrace \widehat{F}_t^{(i)}(T-t), i = 1,\dots,n; t = T-d_{\max}+1,\dots,T-1 \rbrace$. This set is used to compute the simulated nowcasts $\widehat{Y}_t^{(i)} = C_{T-t}/\widehat{F}_t^{(i)}(T-t)$ applying \eqref{eq:nowcast-1}, where $C_{T-t}$ is the observed partial cumulated sum of deaths at time point $T-t$. The point-wise lower and upper bounds of the 95\% prediction intervals for the nowcast for $Y_t$ are then given by the 2.5 and the 97.5 quantiles of the set $\lbrace \widehat{Y}_t^{(i)}, i = 1,\dots,n \rbrace$, respectively.

\section{More Results and Model Evaluation}

\subsection{Spatial Effects}

\begin{figure}[t]
	\center 
	\includegraphics[width = 0.6\textwidth]{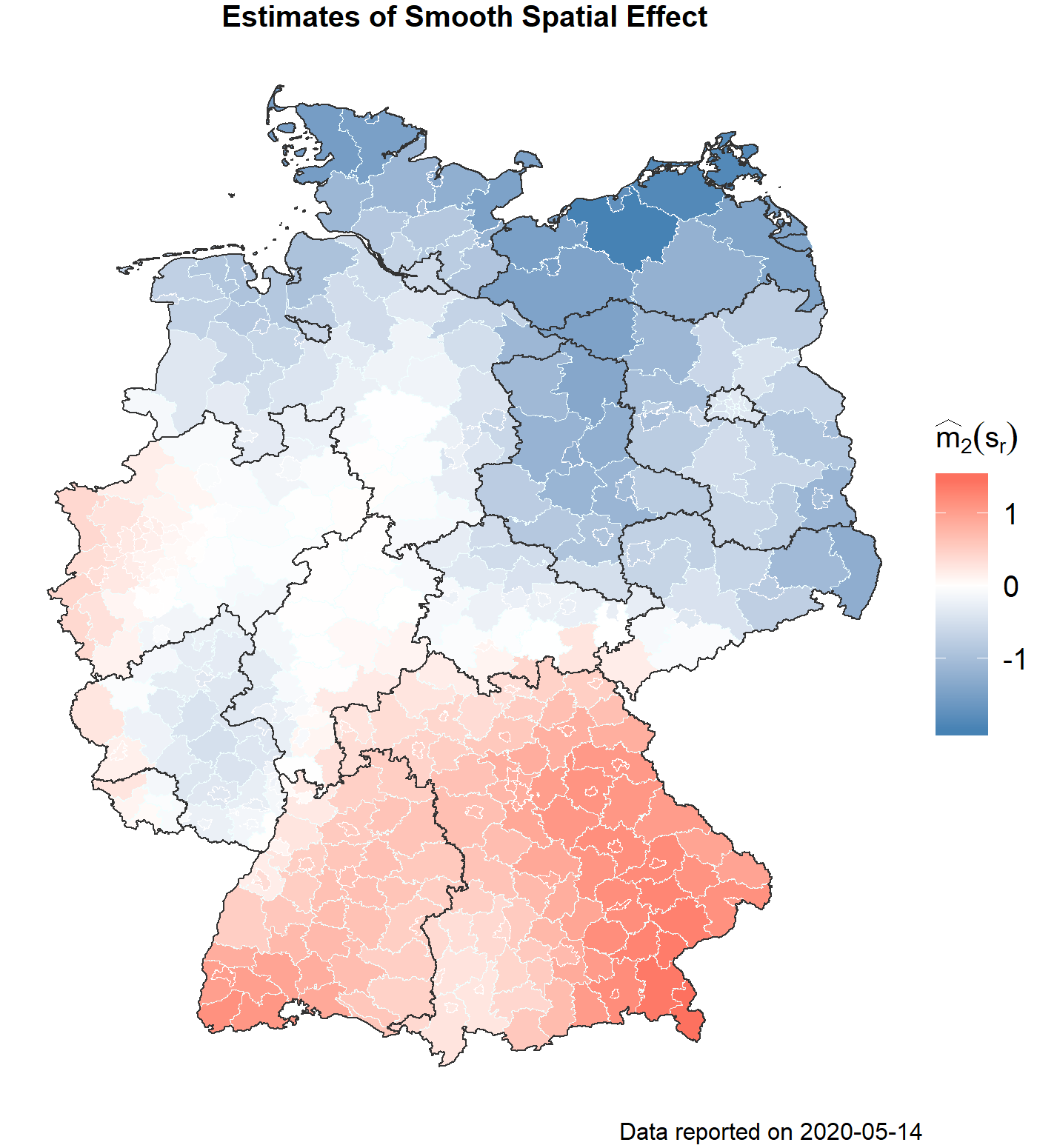} 
	\caption{Smooth spatial effect of the death rate in Germany.}
	\label{fig: spatial effect}
\end{figure}  

\begin{figure}[t]
	\center
	\includegraphics[width = 0.49\textwidth]{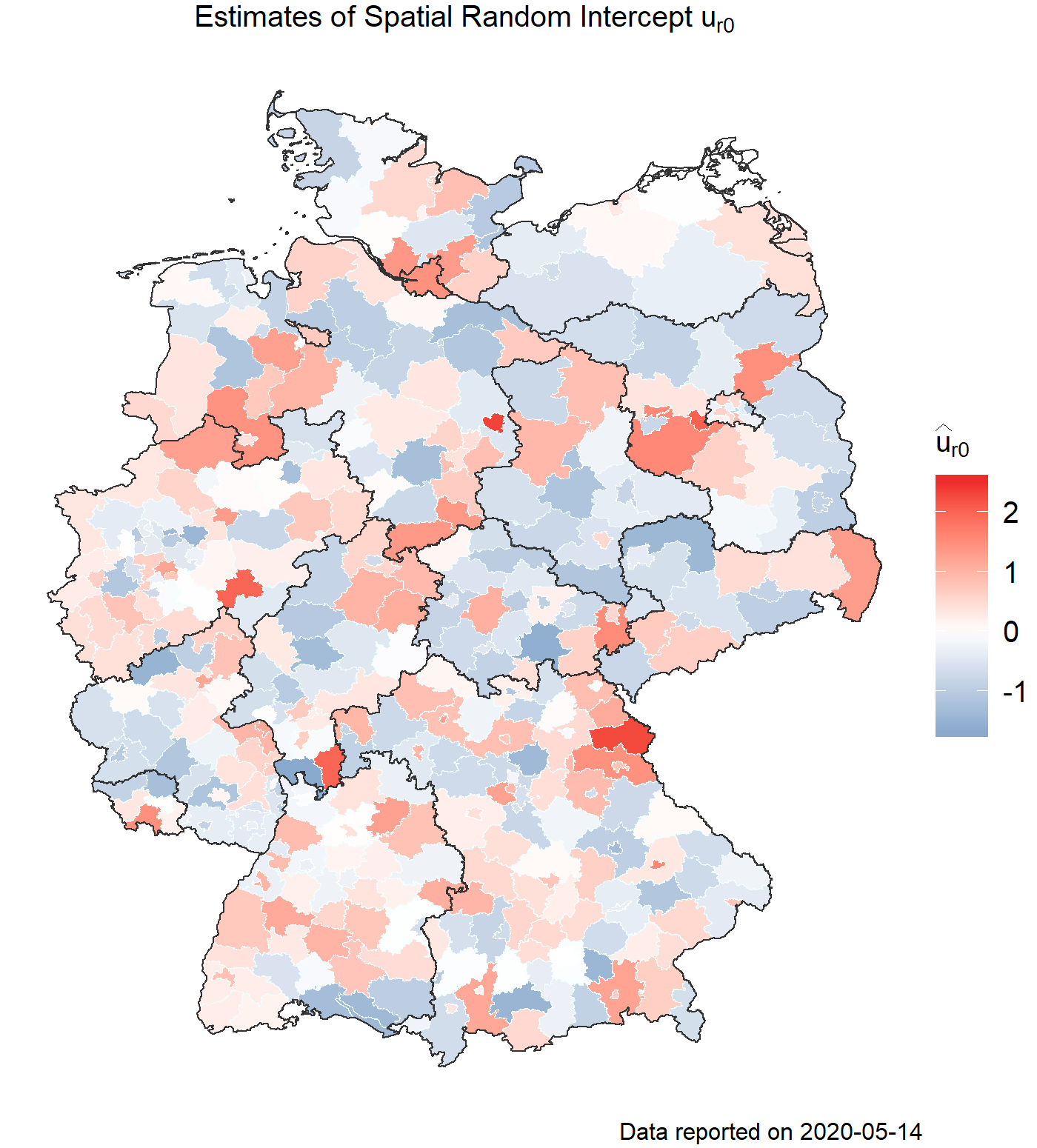} 
	\includegraphics[width = 0.49\textwidth]{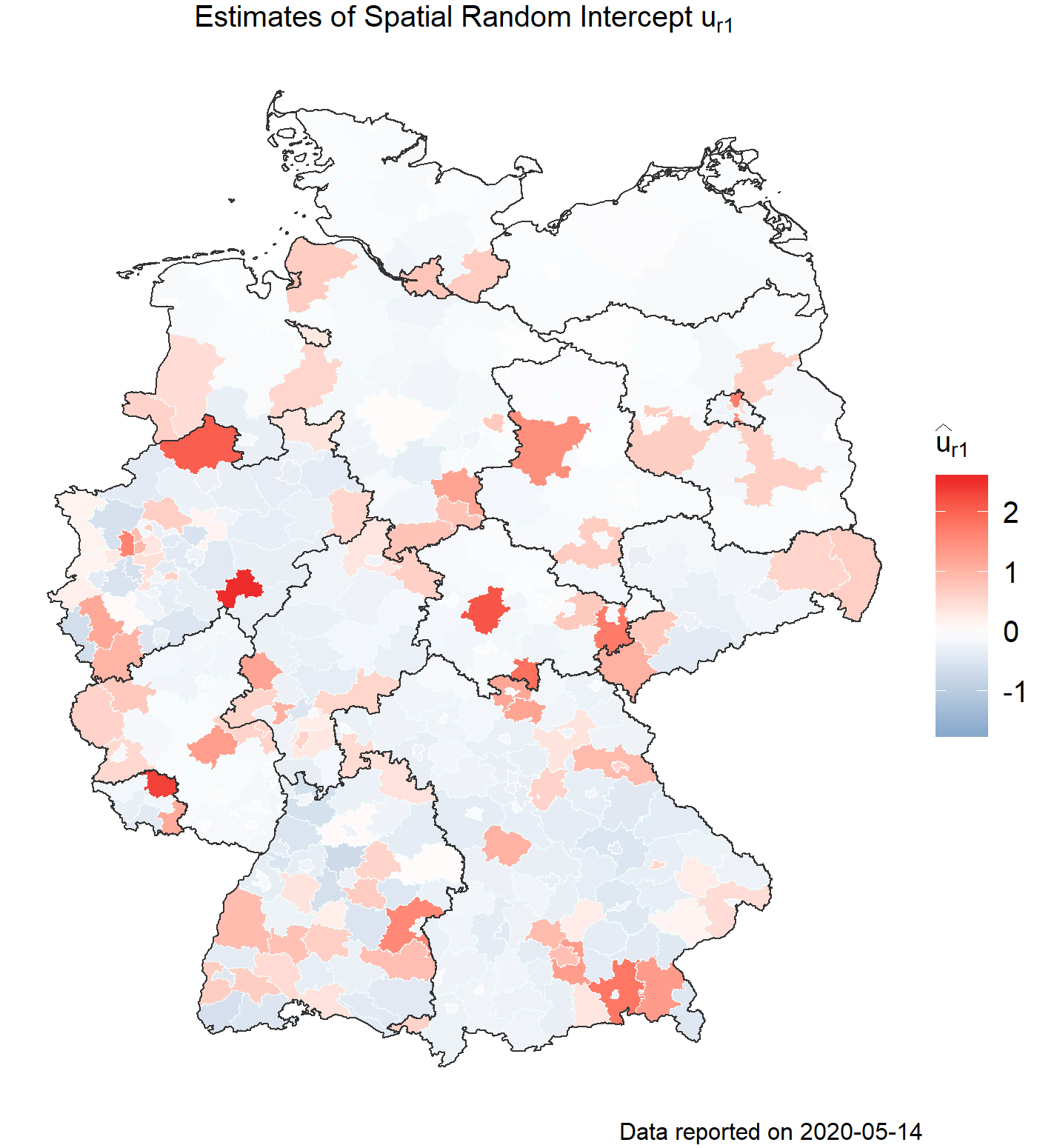} 
	\caption{Long term Region specific level (left hand side) and short term dynamics (right hand side) of the Covid-19 infections} 
	\label{fig: random intercept}
\end{figure}
In Section 3 we presented the fitted death rate, which is the convolution of a  smooth spatial effect as well as region specific effects.  It is of general interest to disentangle these two spatial components. This is provided by the model.  We visualize  the fitted global geographic trend $m_2(\cdot)$ for Germany in Figure  \ref{fig: spatial effect}. The plot confirms that up to May 2020 the northern parts of the country are less affected by the disease in comparison to the southern states. The two plots in Figure \ref{fig: random intercept}  map the region specific effects, i.e. the predicted long term level of a district $u_{r0}$ (left hand side) and  the predicted short term dynamics $u_{r1}$ (right hand side). Both plots uncover quite some region-specific variability. In particular, the short term dynamics captured in the right hand side plot
($u_{r1}$) pinpoint districts with unexpectedly high nowcasted death rates in the last two weeks, after correcting for the global geographic trend and the long term effect of the district. 
Some of the noticeable districts  have already been highlighted in Section 3 above, but we can detect further districts, which are less pronounced in 
Figure \ref{fig: death rate per 100k}. For instance, Steinfurt (in the north-west of North Rhine-Westphalia), Olpe (southern North Rhine-Westphalia) or Gotha (center of Thüringen) presently show a high rate of fatal infections. 

\subsection{Age Group-specific Analyses}
\begin{figure}
	\includegraphics[width = 0.49\textwidth]{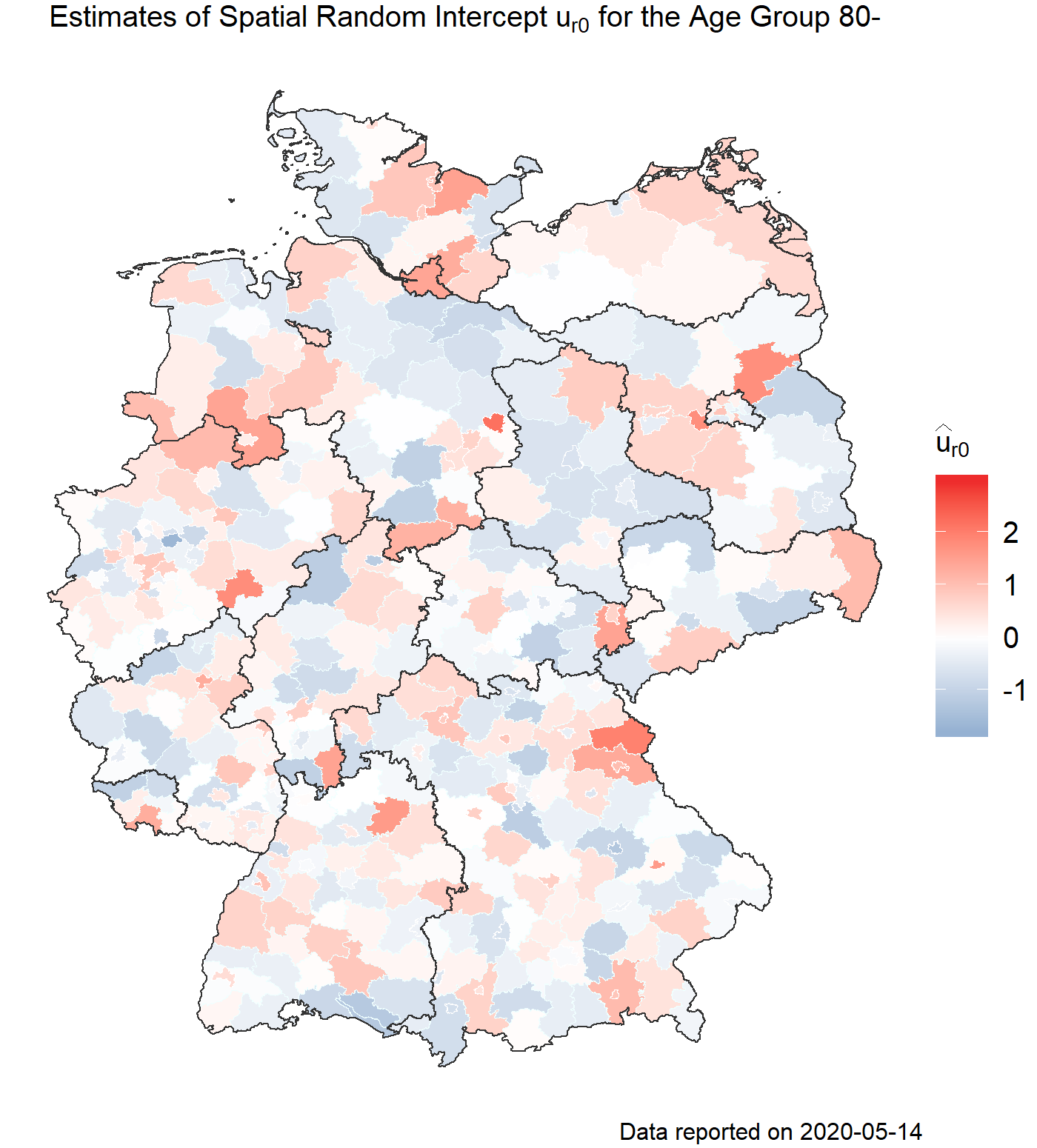} 
	\includegraphics[width = 0.49\textwidth]{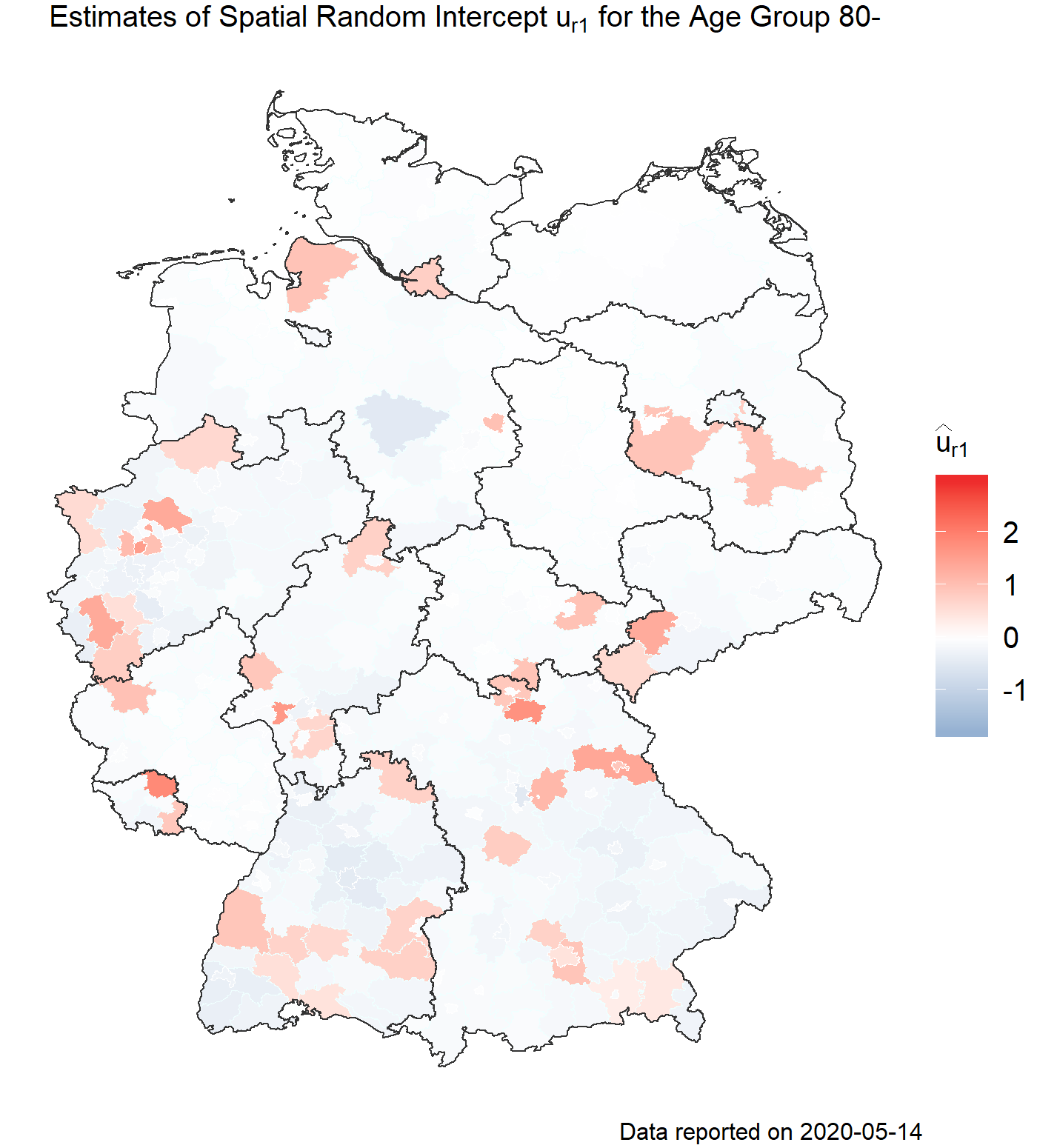} 
	\includegraphics[width = 0.49\textwidth]{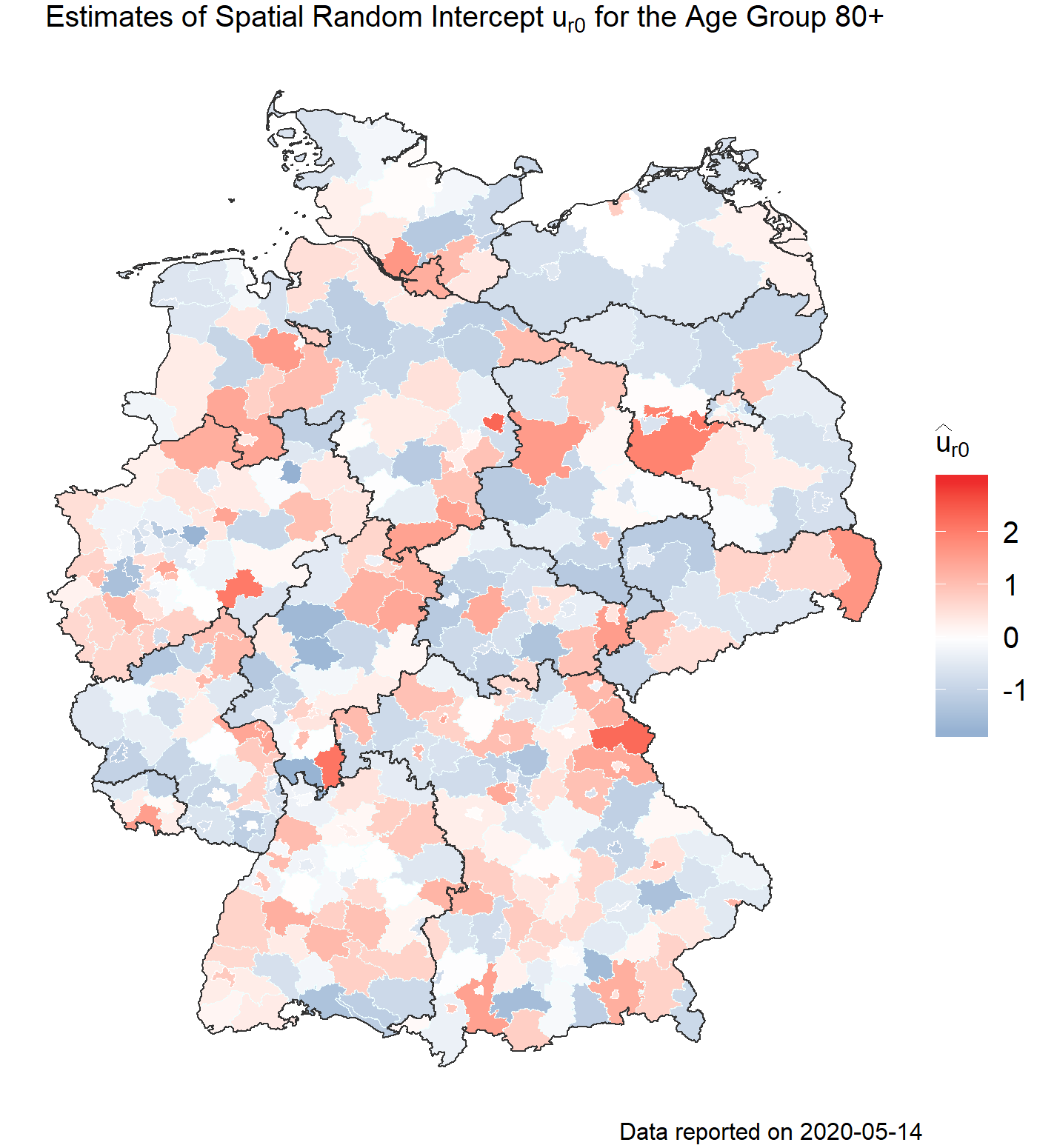} 
	\includegraphics[width = 0.49\textwidth]{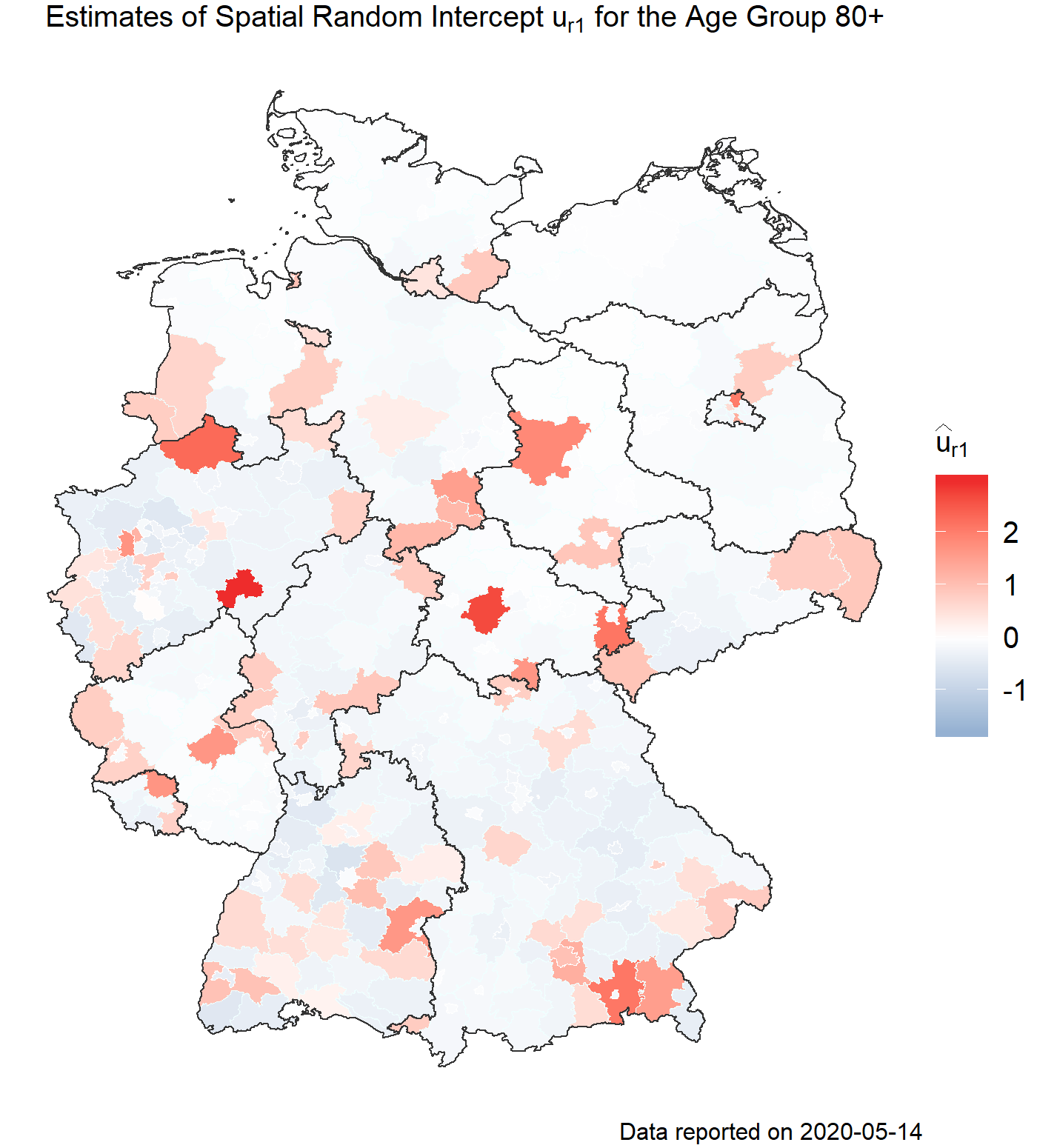} 
	\caption{Region specific level (left hand column) and dynamics (right hand side) of  Covid-19 deaths for the age groups under 80 (80-) and above 80 (80+).}
	\label{fig:separate-age-effects}
\end{figure}

A large number of the registered deaths related to Covid-19 stem from people in the age group 80+.
Locally increased numbers are often caused by an outbreak in a retirement home. Such outbreaks apparently have a different effect on the spread of the disease, and the risk of an epidemic infection caused by outbreaks in this age group is limited. Thus, the death rate of people in the age group 80+ could vary differently across districts when compared to regional peaks in the death rate of the rest of the population. In order to respect this, we decompose the district-specific effects 
$\boldsymbol{u}_r$ in \eqref{eq: lambda rtg} into $\boldsymbol{u}_r^{80-} = (u_{r0}^{80-}, u_{r1}^{80-})^\top$ for the age group 80- and $\boldsymbol{u}_r^{80+} = (u_{r0}^{80+}, u_{r1}^{80+})^\top$ for the age group 80+, where the age group 80- consists of the aggregated age groups 15-34, 35-59 and 60-79. We put the same prior assumption on the random effects as we did in \eqref{eq:u_r}, but now the variance matrix that needs to be estimated from the data has dimension 4 by 4.

The fitted age group-specific random effects are shown in Figure \ref{fig:separate-age-effects}, where the $\boldsymbol{u}_r^{80-}$ are shown in the top panel and the $\boldsymbol{u}_r^{80+}$ in the bottom panel. Most evidently, the variation of the random effects is much higher in the age group 80+ when compared to the younger age groups, as more districts occur  which are coloured dark blue or dark red, respectively. When comparing the district-specific short term dynamics of the last 14 days ($u_{r1}$) in Figure \ref{fig:separate-age-effects} to those in Figure \ref{fig: random intercept}, we recognize that in most of the districts which recently experienced very high death intensities (with respect to the whole period of analysis), these stem from the age group 80+. As mentioned before, this can often be explained by outbreaks in retirement homes. %An exception is here the district Potsdam-Mittelmark which is located southwestward from Berlin and shows a high number of current fatal infection for the population below 80. 

\subsection{Additional Uncertainty in the Poisson Model through the Nowcast}
\label{sec:6.4}
%\begin{figure}[t]
%\center
%\includegraphics[width = 0.8\textwidth]{../../Plots/Nowcasting/OverlayTimeEffect/2020-05-08.pdf} 
%\caption{\textcolor{red}{Fitted smooth death rate per 100 000 inhabitants in the reference group (males aged between 35 and 59 in an average district) including confidence bands adjusted for nowcasting.??}}
%\label{fig: overlay m_1(t)}
%\end{figure}
When fitting the mortality model  (\ref{eq:modquasi}) we included  the fitted nowcast model  as  offset parameter. This apparently neglects the estimation variability in the nowcasting model, which we explored via bootstrap as explained in Section \ref{sec:bootstrap} and visualized in Figure \ref{fig: nowcasting}. In order to also incorporate this uncertainty in the fit of the mortality model, we refitted the model using (a) the upper end and (b) the lower end of the prediction intervals shown in Figure \ref{fig: nowcasting}. It appears that there is little (and hardly any visible) effect on the spatial components, which is therefore not shown here. But the time trend shown in Figure
\ref{fig: death rate per 100k} does change, which is visualized by including the two fitted functions corresponding to the 2.5\% and 97.5\% quantile of the offset function. 
We can see that the estimated uncertainty of the nowcast model mostly affects the last ten days, with a strong potential increase in the death rate mirroring a possible worst case scenario.

\subsection{Residual Analysis in the Nowcasting Model}

\begin{figure}
	\includegraphics[width = \textwidth]{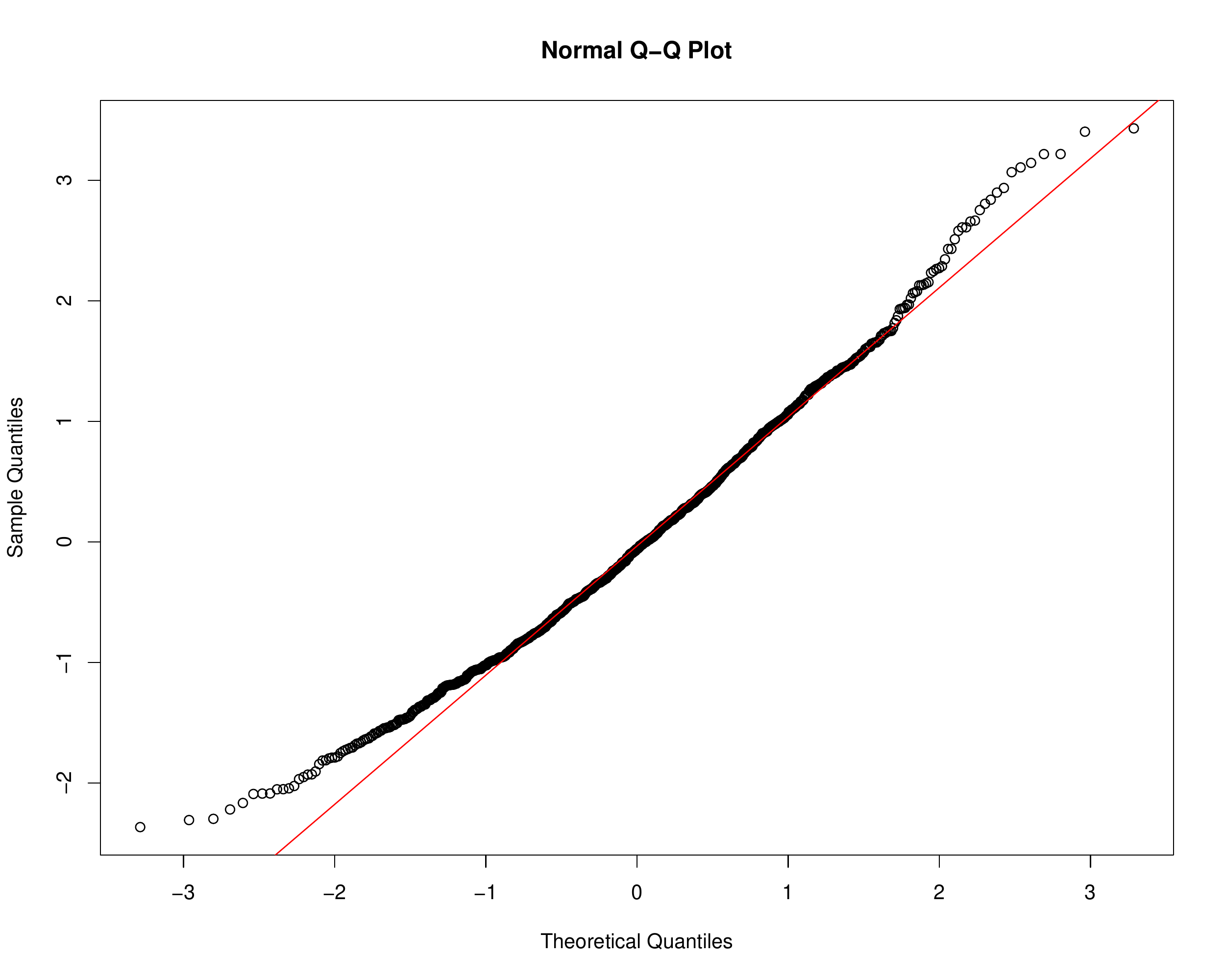}
	\caption{Normal QQ-Plot of Pearson residuals in the nowcasting model.}
	\label{fig: pearson}
\end{figure}

In Figure \ref{fig: pearson} we show a normal QQ-plot of the Pearson residuals in the nowcasting model. Apart from some observations in the lower tail, the Pearson residuals are distributed very closely to a standard normal distribution when considering the estimate $\widehat{\phi} = 1.766$ of the dispersion parameter in the quasi-poisson model \eqref{eq:nowcast-1}. Overall, the model seems to fit to the available data quite well. 

\section{Conclusion}
\subsection{Discussion}
The paper presents a model to monitor the dynamic behaviour of Covid-19 infections based on death counts. It is important to highlight that the proposed model makes no use  of new infection numbers, but only of observed deaths related to Covid-19. This in turn means that the results are less dependent on testing strategies. The nowcasting approach
enables us to estimate the number of deaths following a registered  infection today, even if the fatal outcome has not occurred yet. Moreover, the district level modelling uncovers hotspots, which are salient exclusively through increased death rates. 
A differential analysis of the number of current fatal infections on a
regional level allows to draw conclusions on the current dynamics of the
disease assuming a constant case fatality rate, i.e.\ a
stable proportion of death compared to the true number of infections
when adjusting for age and gender.

A natural next step would now be to consider the nowcasted deaths in relation to the number of newly registered infections, which is, in contrast, highly dependent on both testing strategy and capacity. We consider this as future research, and the proposed model allows us to explore data in this direction. This might ultimately help us in shedding light on the relationship between registered and undetected infections as well as on the effectiveness of different testing strategies.

\subsection{Limitations}

There are several limitations to this study which we want to address as well. First and utmost, even though death counts are, with respect to cases counts, less dependent on testing strategies, they are not completely independent from them. This applies in particular to the handling of post-mortem tests. We therefore do not claim that our analysis of death counts is completely unaffected by
testing strategies. 
Secondly, a fundamental assumption in the model is the independence between the course of the disease and the number of
infections. Overall, if the local  health systems have sufficient capacity and triage can be avoided, this assumption seems plausible, but it is difficult or even impossible to prove the assumption formally. 
Finally,  the nowcasting itself is not carried out on a regional level, though the model focuses on regional aspects of the pandemic. While it would be desirable to fit the nowcast model regionally, the limited amount of data simply prevents us from extending the model in this direction. 

\section*{Acknowledgement}

We want to thank Maximilian Weigert and Andreas Bender for introducing us to the art of producing geographic maps with \texttt{R}. Moreover, we would like to thank all members of the \textbf{Co}rona \textbf{D}ata \textbf{A}nalysis \textbf{G}roup (CoDAG) at LMU Munich for fruitful discussions.

\FloatBarrier

\bibliographystyle{Chicago}
\bibliography{references}

\end{document}